# Phason-mediated interlayer exciton diffusion in $WS_2/WSe_2$ moiré heterostructure

Antonio Rossi*[1,2,8‡], Jonas Zipfel[1,8], Indrajit Maity*[3,8], Monica Lorenzon[1], Luca Francaviglia[1], Emma C. Regan[4], Zuocheng Zhang[4], Jacob H. Nie[4,5], Edward Barnard[1], Kenji Watanabe[6], Takashi Taniguchi[7], Eli Rotenberg[2], Feng Wang[4], Johannes Lischner[3], Archana Raja[1]* and Alexander Weber-Bargioni[1]*

1-The Molecular Foundry, Lawrence Berkeley National Laboratory, Berkeley, USA.
2-Advanced Light Source, Lawrence Berkeley National Laboratory, Berkeley, USA.
3- Imperial College London, South Kensington Campus, London, United Kingdom.
4- Department of Physics, University of California at Berkeley, Berkeley, CA, USA.
5- Department of Physics, University of California at Santa Barbara, Santa Barbara, CA, USA
6- Research Center for Functional Materials, National Institute for Materials Science, 1-1 Namiki, Tsukuba, Japan 8
7-International Center for Materials Nanoarchitectonics, National Institute for Materials Science, 1-1 Namiki, Tsukuba,Japan
8-These authors contributed equally

‡-Current affiliation: Center for Nanotechnology Innovation @ NEST, Instituto Italiano di Tecnologia, Pisa,Italy

*antonio.rossi@iit.it, i.maity@imperial.ac.uk, araja@lbl.gov, awb@lbl.gov,

**ABSTRACT**

Moiré potentials in two dimensional materials have been proven to be of fundamental importance to fully understand the electronic structure of van der Waals heterostructures, from superconductivity to correlated excitonic states. However, understanding how the moiré phonons, so-called phasons, affect the properties of the system still remains an uncharted territory. In this work, we demonstrate how phasons are integral to properly describe and understand low temperature interlayer exciton diffusion in $WS_2/WSe_2$ heterostructure. We perform photoluminescence (PL) spectroscopy to understand how the coupling between the layers, affected by their relative orientation, impacts on the excitonic properties of the system. Samples fabricated with stacking angle of 0° and 60° are investigated taking into account the stacking angle dependence of the two common moiré potential profiles. Additionally, we present spatially and time-resolved exciton diffusion measurements, looking at the photoluminescence emission in a temperature range from 30 K to 250 K. An accurate potential for the two configurations is computed via density functional theory (DFT) calculations. Finally, we perform molecular dynamics simulation in order to visualize the phasons' motion, estimating the phason speed at different temperatures, providing novel inside into the mechanics of exciton propagation at low temperatures that cannot be explained within the frame of classical exciton diffusion alone

**INTRODUCTION**

Van der Waals (vdW) heterostructures are often considered key to unlock the full potential of two-dimensional (2D) materials[1–3]. Weak interaction at the interface makes it possible to virtually stack any 2D system on top of another, creating artificial heterostructures with entirely novel properties. Twisted bilayer graphene (TBG), for example, displays flat bands near the Fermi level[4–6]. The heavy fermions populating these states give rise to unconventional superconductivity[7,8], but also insulating behaviour driven by electron correlations[9]. Strong electron correlation is also observed in transition metal dichalcogenide (TMD) heterostructures like $WSe_2/WS_2$, where flat bands arise[10,11] and a Mott insulator state and generalized Wigner crystallization are observed[12]. Some of the new properties observed and predicted in 2D vdW heterostructures[13,14] are determined by the influence of their moiré potential, the super periodicity arising from the superposition of two layers with different lattice constant and/or relative orientation. Different lattice parameters and/or stacking angles give rise to the formation of a moiré lattice that translates into a potential landscape[15,16]. Therefore, such a heterostructure is not simply the result of the sum of the constituents, but new unexpected phenomena can emerge. An unsettled question is whether the moiré potential is stationary or dynamic at finite temperature as the dynamic moiré potential can significantly influence the observed phenomena. Here, we tackle this challenge by employing the temperature dependent propagation of interlayer excitons (IX) in $WS_2/WSe_2$ as probes for the moiré potential dynamics. Our experiments establish that the moiré potential is dynamic at finite temperature and the quasiparticle describing its motion, the phason[17,18], strongly affects the exciton diffusion mechanism at low temperatures.

The $WSe_2/WS_2$ system is made up of stacked direct band gap monolayers and can host both intralayer exciton and IX[19,20]. The latter is formed when the electron and the hole reside in different layers. The IX's radiative lifetime is extremely long[21,22] and can reach hundreds of nano-seconds in some systems[23]. Its spectroscopic and dynamical features are dramatically affected by the presence of a moiré potential, as opposed to the exciton diffusion in a single monolayer where such moiré modulation is not present[24,25]. A first study reporting the absorption spectroscopy on the IX dynamics and diffusion on a CVD grown $WSe_2/WS_2$ system has been reported by Yuan et al[26]. Furthermore, a trapping/de-trapping process, which is a temperature activated mechanism where thermal energy allows the exciton to hop across the moiré potential wells, has previously been proposed for a $MoSe_2/WSe_2$ heterostructure at various twist angles[27]. However, we note that our observations of IX diffusion cannot be explained by such a model alone. For that reason we take into account the role of phonons, specifically of moiré-induced phonons also known as phasons[17], that are still experimentally unexplored. Recent calculations demonstrate how these quasi-particles can play a crucial role in understanding the charge ordering and correlated

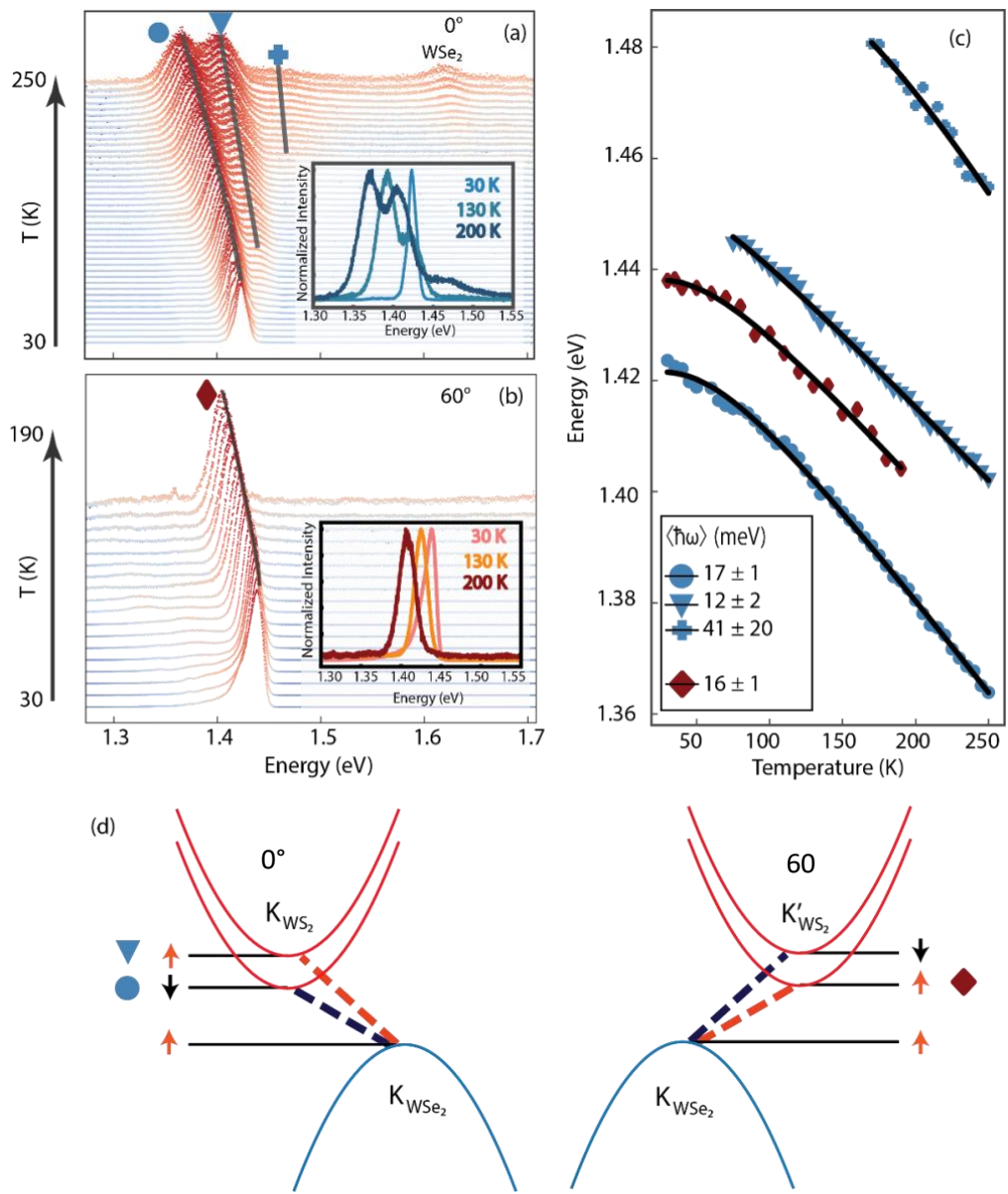

Figure 1 PL spectra of 0° (a) and 60° (b) stacking angles at different temperature. Insets display representative spectra collected at different temperatures. Blue and red symbols identify the different peaks present in the spectra. In panel (a) an WSe$_2$ intralayer exciton peak is highlighted at higher temperatures. (c) Peak position tracked as a function of temperature fitted with Varshni equation (black line). Inset displays the value of the average phonon energy for each fit. (d) Scheme of the spin-polarized band structure for the 0° and 60° (left and right panels, respectively) structures with the corresponding vertical optical transitions (dashed lines).

states of magic-angle TBG[28,29].

In order to gain a full understanding of the role of moiré phonons in IX dynamics, we perform a wide-range temperature dependence study of the IX diffusion through direct photoluminescence (PL) detection. We explore the PL spectra of the interlayer excitons as a function of temperature and stacking angle (0° and 60°), together with the lifetime and diffusion length/coefficient[30]. This allows us to gain deep insight, first, into the temperature dependent spectral properties of the IX for the two stacking configurations with different moiré landscape, and then into the IX propagation dynamics. Our findings go beyond the classical Arrhenius model of thermally activated exciton transport involving a simple fixed potential landscape that have been reported in previous works where exciton diffusion is negligible at low temperatures (below 100 K)[26,27,31]. Instead, we find a regime between 30 K and 100 K where excitons can explore the system, despite being fully trapped inside deep moiré potentials. Finally, a model that describes the electronic properties and vibrational modes of the two systems is proposed, showing how the moiré potential itself is affected by temperature and is likely the main mechanism for a surprising non-zero diffusion coefficient at cryogenic temperatures, where excitons are fully trapped. An accurate model describing the moiré potential landscape is also provided, matching the energy barrier obtained from temperature-dependent diffusion measurements.

**Results & Discussion**

The different twist angles affect the electronic properties of the heterostructure, dictating the alignments of spin-valley locked bands. The IX is an electron-hole pair where the two charged particles are located in different layers and the recombination is affected by the selection rules dictated by the different twist angle. To visualize the IX spectral properties, we analyse the PL emission as a function of temperature. Figure 1 (a,b) reports the PL spectra of the two $WS_2/WSe_2$ heterostructures with 0° and the 60° alignment in blue and orange, respectively. The PL spectra are reported at increasing temperatures, starting from 30 K. For both stackings, the emission is entirely dominated by the IX involving the top of the $WSe_2$ valence band and the bottom of the $WS_2$ conduction band at K and K'. Such emission is centered at 1.42 eV for the 0° stacking and 1.44 eV for the 60° stacking at 30 K, and progressively red shifts with increasing temperature, as expected[32,33]. Interestingly, for the 0° structure two thermally activated IXs emerge: a first one at around 75 K, and a second one above 150 K. The insets of Fig 1(a,b) show representative spectra featuring all peaks at three selected temperatures. We can describe the peaks evolution by tracking their position as a function of temperature, as displayed in Fig 1c. We evaluate each peak position by fitting the spectra with either one, two or three Voigt functions, depending on the number of observed peaks (see supporting information Fig. S2.). Using Varshni's equation, which empirically describes the temperature dependence of energy gaps in semiconductors[31], we extract an empirical electron-phonon coupling coefficient associated with the average phonon energy $\langle \hbar \omega \rangle$. The fitted parameters values are reported in the inset of panel (c). The two lowest excitons from the 0° structure (marked with blue circles and triangles (Fig. 1a) have similar fitting parameters, suggesting very similar electron-phonon coupling. Moreover, they display an energy separation of $\Delta E \sim 30$ meV, compatible with the spin orbit splitting of the conduction band in $WS_2$[34]. The origin of the third, highest energy peak marked by a blue cross, which displays a much steeper slope, is unclear as of now; however, it could be associated with one of the moiré excitons previously observed in the same kind of heterostructure for the intralayer exciton[14]. Another plausible hypothesis takes into account the description of the moiré potential as a quantum well; within this hypothesis, the third peak arises from the optical transition involving the first excited state of the well. The PL spectra collected for the 60° stacking sample show a different behaviour. There is only one peak (marked with red diamond) and it shows an identical trend as the two lowest emission peaks from the 0°sample. The presence of three peaks in the 0° alignment and the single peak emission for the 60° can be ascribed to the selection rules coming from the electronic structure of the single layers[35,36]. The lowest energetic transition for the 0° sample should be related to the spin dark exciton. However, the presence of the moiré potential has been demonstrated to partially lift those selection rules in systems with large lattice mismatch[37] and, given it is the lowest energy state, its high population at low temperatures makes the PL emission comparably strong, similarly to that observed for $MoSe_2/WSe_2$ structure[38]. With increasing temperature, however, the tail of the Fermi-Dirac distribution starts populating the higher spin-split level of the $WS_2$ conduction band, allowing the spin-aligned IXs to become observable[20]. On the other hand, the 60° sample has a bright transition as the lowest available one, allowing all the hot electrons to relax and recombine through that channel, while the higher spin-dark state, being much less efficient and much less populated even for elevated temperatures, is not easily observed (Fig 1(d)).

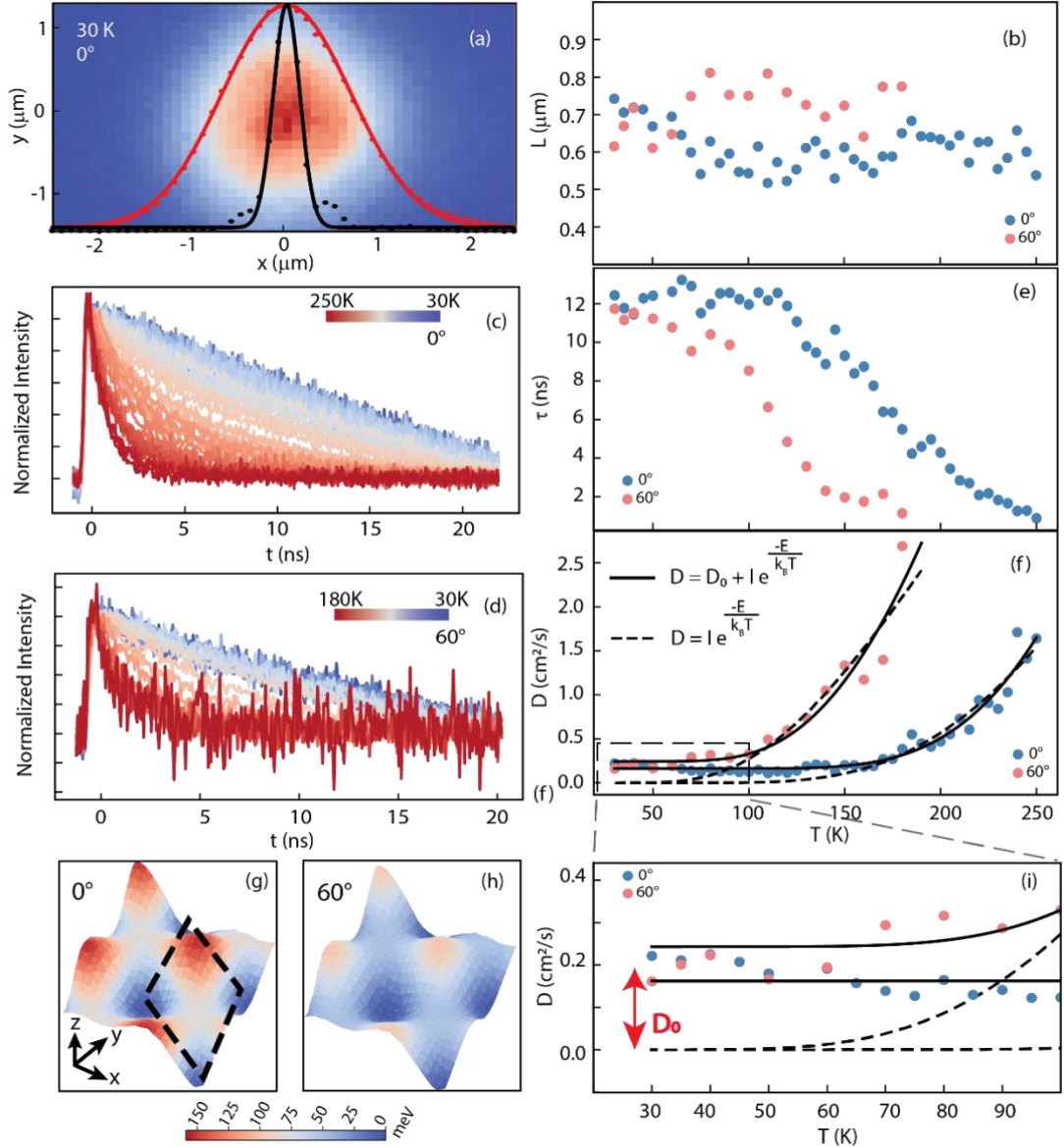

*Figure 2 (a) Exciton diffusion length evaluated from PL area with respect the laser excitation width at 30K. (b) Diffusion length values at different temperature for the two stacking angles. PL lifetime traces at different temperature for 0° (c) and 60° (d) stacking angle.(e) PL recombination time as a function of temperature for the two stacking angles. (f) Diffusivity values fitted with (solid line) and without (dashed line) $D_0$ offset. (g) and (h) moiré potential evaluated as described in the main text for 0° and 60° respectively. The dashed black diamond indicates the moiré unit cell of periodicity $\lambda \sim 7nm$. (i) Zoom-in of the diffusivity map at a lower temperature range to highlight the $D_0$.*

We now focus our attention on the IX dynamics, studying its diffusion and recombination lifetime. We perform an exciton diffusion study by collecting the light associated with the IX transitions only and mapping it in real space (supporting information Fig. S(3-4)). To achieve this goal, we use a 790nm long pass filter in detection to cut out any potential light emitted from the intralayer transitions, particularly at elevated temperatures.

The diffusion length is defined as[39]

$$L = \sqrt{\sigma^2 - \sigma_L^2}, \qquad \text{Eq. 2}$$

where $\sigma_L$ is the gaussian covariance that fits the centrosymmetric excitation laser profile along an arbitrary axis (Fig 2(a)) and $\sigma$ is the gaussian covariance of the real space PL intensity, evaluated in the same manner, assuming that, if excitons can diffuse, the emission of the initially excited 2D-gaussian profile broadens isotropically over time, following Fick's second law. The diffusion length $L$ is evaluated as the square root of the difference between the squared covariances, according to Eq. 1 (Fig 2(b)).

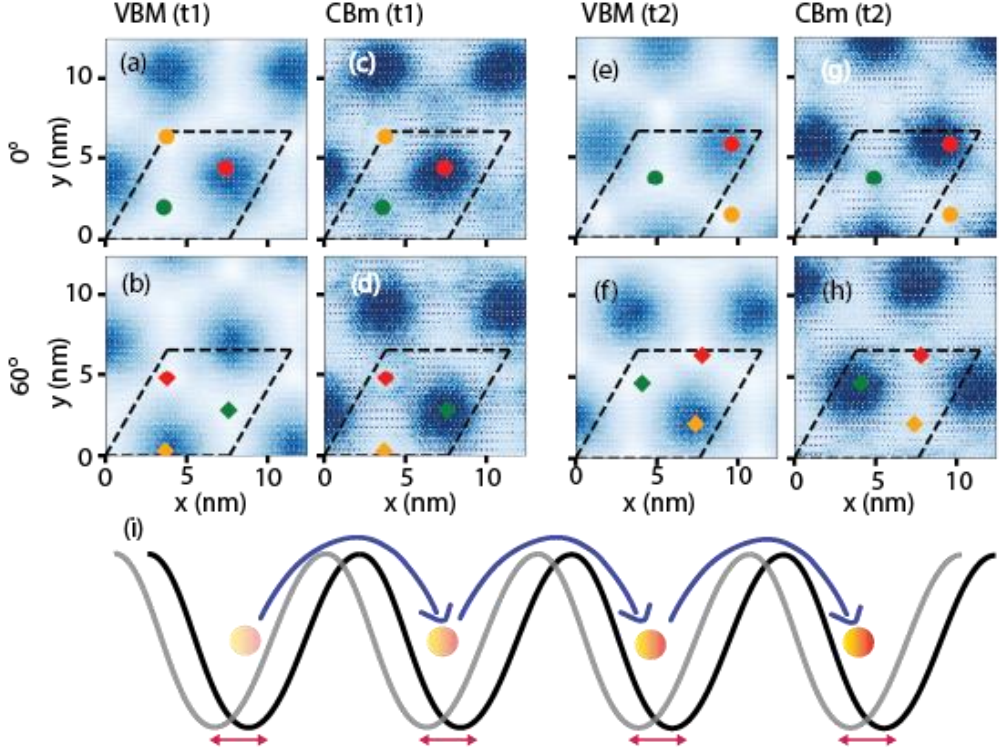

*Figure 3 Local holes and electrons densities evaluated at valence band maxima (VBM) and conduction band minima (CBM) respectively for 0° and 60° deg structure. VBM and CBM densities are evaluated at different times, t1 (a-d) and t2(e-h) for finite temperature T=175K with t2-t1=0.3 ns. The colored dots and diamonds track the different moiré sites of the heterostructure. (i) Scheme of the exciton (yellow sphere) hopping assisted by moiré phasons.*

Simultaneously to the spatially resolved diffusion measurements, we collect the time resolved PL emission. Fig 2(c,d) show data for the 0° and 60° samples respectively. The decay traces exhibit a single exponential trend, which we fit to extract the PL lifetime $\tau$ (Fig 2(e)).

The diffusion length for both stacking angles is largely unaffected by temperature variations. It is mostly confined within a mean value of ~ 600 nm, which is consistent with reported diffusion lengths in monolayer systems[40,41]. On the other hand, the PL lifetime shows a clear decreasing trend with increasing temperature, dropping after 150 K for 0° and 100 K for 60°. Using these values of L and $\tau$ we extract the diffusion coefficient, defined as[40,42]

$$D = \frac{L^2}{2\tau} \qquad \text{Eq. 3}$$

Fig 2(f) shows the calculated values of the diffusion coefficient D (diffusivity), which undergoes a thermally activated process resulting in a pronounced increase of the diffusivity with increasing temperature. Such an exponential behaviour expected when considering a simple moiré-trap picture. Once formed, the IXs are affected by the presence of the moiré potential landscape and are trapped in its local minima. However, if enough thermal energy is provided and the activation barrier is overcome, the IX can now diffuse faster, reaching a value comparable with the one of intralayer excitons of an isolated monolayer where the moiré potential is not present (see supporting information Fig. S5). To evaluate the effective potential for diffusion, we perform extensive density functional theory (DFT) calculations for both 0° and 60° twisted heterobilayers, taking into account the atomic reconstructions at T = 0 (see supporting information). The moiré potential is evaluated as

$$\Delta_M(\boldsymbol{r}) = E_g(\boldsymbol{r}) - E_g^{min}. \qquad \text{Eq. 4}$$

$E_g(r)$ represents the local bandgap and $E_g^{min}$ represents the averaged bandgap[35]. Since $\Delta_M(r)$ varies smoothly over the moiré period, the effective moiré potential for diffusion can be extracted from local bandgap at different high-symmetry stackings. The averaged effective moiré potential for 0° (124 meV) is unmistakably larger than that of 60° $WS_2/WSe_2$ (85 meV) (Fig2 (g,h)).

The most striking observation however is that a simple detrapping model, however, is not consistent with what observed at low temperatures (T < 100 K). Fig 2 (f) displays the diffusivity data reported in the dashed boxed in Fig 2 (e). We emphasize that an Arrhenius-like description would make D tend to 0 at low temperatures (dashed line). However, the data clearly show a finite offset $D_0$, indicating that the IXs are diffusing even at low temperatures. Having excluded any relevant quantum effects due to the low exciton density (see Experimental methods section), we propose moiré phonons (phasons)[17] to be responsible for this deviation from the Arrhenius description. The phason modes represent an effective translation of the moiré sites due to the non-uniform out-of-phase in-plane translation of two constituent layers in the heterobilayer. We demonstrate that the phason modes give rise to dynamic moiré potentials at finite temperature. We simulate a $WS_2/WSe_2$ heterostructure in both twist angle configurations, in a canonical ensemble using classical molecular dynamics simulations. We find that the moiré pattern starts to move in-plane due to the time dependent relative local displacement of the two layers as a result of the thermally activated phasons. Figure 3 (a-h) shows different temporal snapshots of the valence band maximum (VBM) and conduction band minimum (CBM) single particle density of states for 0° and 60° stacking at a lattice temperature of 175 K. The molecular dynamics simulation reveals, as evidenced by the coloured marks, that the different stacking sites, and thus the moiré potential, move with time. It is also shown that the single particle density of states calculated from functional density theory strictly follow this movement as would be expected within the Born-Oppenheimer approximation (See supporting information Fig S. 8. An animated simulation is found in the online supporting information).

Glazov and coworkers[43] predicted how the exciton-phonon interaction can lead to a finite diffusivity term of the form

$$D_0 = \frac{v^2 \rho \hbar^2}{m^2 E_D^2} \qquad \text{Eq. 5}$$

where v is the speed of sound, ρ is the mass-density, m is exciton mass, and $E_D$ is the deformation potential. Based on Eq.5 we estimate the diffusivity at low T to be ≈2-3 cm$^2$/sec. using previously reported parameters of monolayer TMDs[44]; this value is independent of T. In this model we assume linear dispersion for acoustic phonons (LA phonon) and parabolic dispersion for excitons. In our experiments, the diffusivity at low-T is at least one order of magnitude lower than the diffusivity predicted in Eq. 5 for a monolayer. This may be attributed to the lower speed of phasons (modes P1 in Fig 4(a,b)) than a single layer phonon. Also, an increase in the exciton mass and a deformation potential arising from exciton-phonon coupling have to be taken into account. It is worth emphasizing that Eq.5 is oblivious to the moiré potential being dynamic but incorporates small-wavelength phonons. On the other hand, the dynamic moiré potential discussed earlier arises only from the very-long wavelength phasons, allowed by the simulation cell of length $L_{sim}$ with $q = 2\pi/L_{sim}$. The dynamic moiré potential effectively changes the speed of sound for very-long wavelength phasons in Eq. 5 and thus, impacts diffusivity.

The diffusivity as a function of temperature now takes the form

$$D = D_0 + I e^{\frac{-E_0}{k_B T}} \qquad \text{Eq. 6}$$

The moiré potential height E0 can therefore be extracted from the data resulting in a value of 126 meV and 67 meV for 0° and 60° orientation, respectively, in great agreement with the calculated potential shown in Fig 2 (g,h). A value of $D_0$ = 0.16 cm²/s and 0.24 cm²/s is also obtained from the fitting for 0° and 60° respectively, allowing for a first indirect measurement of exciton-phason coupling.

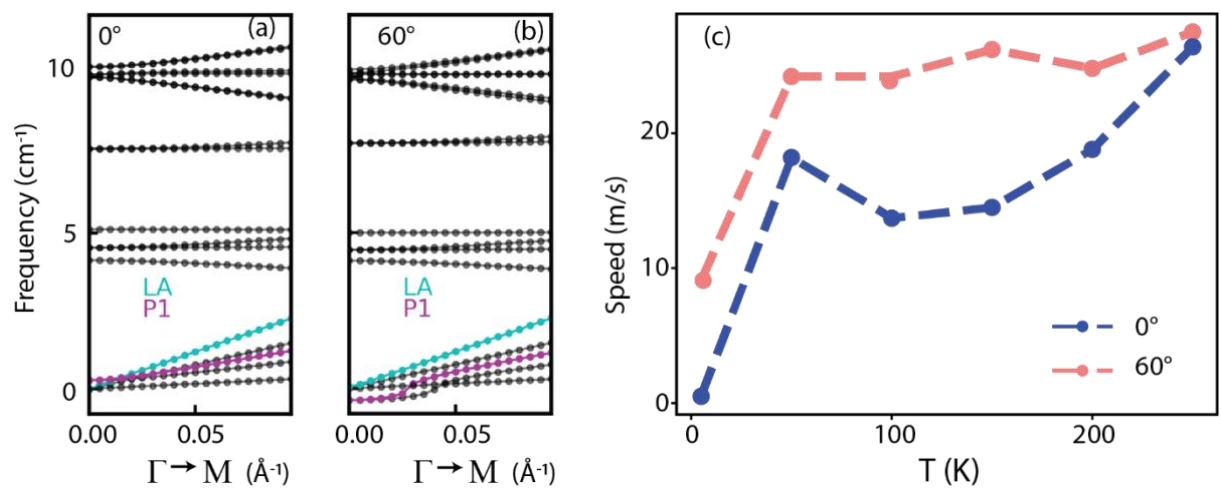

*Figure 4 (a,b) Phason dispersion overlayed to the phonon dispersion for 0° and 60°, respectively computed at T=0 K for the long-wavelength phonons near the Γ point. The phason speed is reduced by a factor of 2 and 1.6 when compared to the LA mode for 0° and 60° twist angles, respectively (c) Speed of moiré sites captured using molecular dynamics simulations as a function of temperature for 0° and 60°.*

Finally, it is possible to extract the speed of moiré sites as a function of temperature by evaluating the stacking site displacement at different times (Fig. 4 (c)). This temperature dependence has a very steep incline and then quickly saturates, in agreement with the expectation that the sliding of two dissimilar layers costs almost no energy, which is generally referred to as super lubricant behavior[45,46].

The calculated speed of the moiré pattern is about a factor of 3 lower than what would be expected from our measurements, simply obtained by dividing the diffusion length by the recombination lifetime. The proposed dragging of excitons by a phason-driven moving of the moiré pattern still provides as an interesting explanation for the observed non-zero diffusion at cryogenic temperatures, particularly because this behavior should be expected for any moiré system. We also note that for a more detailed analysis of these phason driven moiré dynamics, a more evolved model, beyond the classical scope of the here used molecular dynamics should be taken into consideration. However, regardless of the relative contributions of phason-driven dynamic moiré potential and of the exciton-phason scattering as in Eqn.5 to total diffusivity at low-T, our calculations strongly indicate phasons play a key role in exciton dynamics.

**Conclusion**

In summary, we study the excitonic properties of $WSe_2/WS_2$ at two different stacking angles as a function of temperature. We observe a different PL signature depending on the intrinsic selection rules of the monolayers forming the heterostructures. Exciton diffusion measurements offer a way to probe the moiré barrier, estimating a shallower 67 meV potential for the 60° compared to a 126 meV potential of the 0° stacking. Most strikingly, we observe a non-zero diffusion at cryogenic temperatures reflected in a constant offset of the diffusion coefficient that cannot easily be explained within the frame of a de-trapping model. To elucidate this behavior, we propose the mechanism of excitons being dragged across the crystal by a moving moiré potential landscape, that could already occur from temperatures close to 5K, as estimated from our molecular dynamics simulations, and that should generally be applicable to any moiré system. In conclusion, exciton properties are deeply affected by the moiré dynamics and IXs themselves can be used as probes to investigate the moiré motion.

**Experimental Methods**

We start by preparing the heterostructure with a standard pick up technique[47]. The TMD heterostructure is incapsulated in h-BN to preserve the pristine properties of the system. The relative alignment has been checked using second harmonic generation (supporting information Fig. S1). The experiment has been carried out using a Montana Cryostation s200 closed-cycle helium cryostat with full temperature control down to 4.2K and 100x optical magnification (NA=0.95). The samples are excited using a pulsed (38.9 MHz) supercontinuum laser tuned to 510 nm with 10 nm linewidth and 30ps pulse length, focused down to a spot diameter of about 320 nm FWHM, yielding an energy density per pulse of 0.38 µJ/cm² for our diffusion experiment. Assuming 5% absorption mainly from the $WS_2$ B-exciton and unity conversion into interlayer excitons, this amounts to an exciton density of roughly $5\times10^{10} cm^{-2}$, within a regime where we can exclude a non-linear response. At this density, quantum effects due to dipole-dipole interaction are negligibly small compared to the thermal energy in the range of temperatures explored[26]. PL spectra are collected using a spectrometer with dispersion grating together with a Peltier-cooled charge-coupled device (CCD) camera while lifetimes are measured using an avalanche photo diode (APD) with a nominal temporal resolution of 30ps. Real space imaging of the PL emission to extract exciton diffusion lengths is realized using a fast and highly efficient CMOS detector in time integrated detection mode. We acquired Raman spectra using an ultra-narrow diode pumped solid state (DPSS) laser at 532 nm focused into a diffraction-limited spot by means of a 40X objective. We set the power output at the objective end to 10 µW, corresponding to a power density of about 1000 W/cm². The scattered light was collected onto a Peltier-cooled CCD through a spectrometer with a 1800 lines/mm grating and a notch filter to reject the laser line. We integrated each spectrum for 20 minutes. For low-temperature measurements, the sample was place in a flow-through cryostat operated with either liquid Helium or liquid Nitrogen.

**Computational Methods**

Twisted $WS_2$/$WSe_2$ heterobilayers have been generated using the TWISTER package[48]. All the structural relaxations, molecular dynamics, and phonon calculations are performed using classical interatomic potentials fitted to density functional theory calculations. The intralayer interactions within $WSe_2$ and $WS_2$ are described using Stillinger- Weber potential[49], and the interlayer interactions are captured using the Kolmogorov-Crespi potential4. All the simulations with classical interatomic potentials are performed using the LAMMPS package[50,51] and the phonon calculations are performed using a modified version of the PHONOPY package[52]. We relax the atoms within a fixed simulation box with the force tolerance of 10−6 eV/Å for any atom along any direction. All the molecular dynamics simulations are performed in the canonical ensembles for several temperatures and the moiré movements are extracted in the microcanonical ensemble. The speed of moiré sites is extracted from molecular dynamics simulations of a 3×3×1 supercell of the moiré unit-cell. We find that the moiré sites move even for a simulation of a 20×20×1 supercell. Note that the supercell size dictates the lowest accessible momentum close to Γ point, i.e., $q_{min} \propto 1 L_m$, where $L_m$ is the moiré simulation cell lattice constant. The moiré site speed differences between twist angles 0° and 60° are reproduced for 1×1×1 and 6×6×1 supercell, as well. All the electronic structure calculations are performed using the SIESTA package[53]. All the calculations include spin-orbit coupling[54]. We have used local density approximation with the Perdew-Zunger parametrization as the exchange-correlation functional[55].

**Acknowledgements**

The authors want to thank Prof Libai Huang from Purdue University, Prof Alexey Chernikov from Technische Universität Dresden, Dr. Mit H Naik from University of California Berkeley, Dr. Elyse Barré and Dr. Medha Dandu from The Lawrence Berkeley National Lab for fruitful discussion. This research was supported as part of the Center for Novel Pathways to Quantum Coherence in Materials, an Energy Frontier Research Center funded by the US Department of

Energy, Office of Science, Basic Energy Sciences. The experimental work was performed at the Molecular Foundry, which was supported by the Office of Science, Office of Basic Energy Sciences, of the U.S. Department of Energy under contract no. DE-AC02-05CH11231. I.M. acknowledges funding from the European Union's Horizon 2020 research and innovation program under the Marie Sklodowska-Curie Grant agreement No. 101028468. This work used the ARCHER2 UK National Supercomputing Service (https://www.archer2.ac.uk).

## Code Availability

The twisted heterobilayer structure construction, atomic relaxations, molecular dynamics simulations, phonon spectra calculations, and electronic band structure calculations presented in the paper were carried out using publicly available codes. Our findings can be fully reproduced using these codes. Some of the post-processing tools, such as tracking the motion of moiré sites, developed for the paper will be made publicly available at [https://gitlab.com/_imaity_/Moiredynamics](https://gitlab.com/_imaity_/Moiredynamics).

# Supporting Information: Phason-mediated exciton diffusion in WS$_2$/WSe$_2$ moiré heterostructure

Antonio Rossi*[1,2,8‡], Jonas Zipfel[1,8], Indrajit Maity*[3,8], Monica Lorenzon[1], Luca Francaviglia[1], Emma C. Regan[4], Zuocheng Zhang[4], Jacob H. Nie[4,5], Edward Barnard[1], Kenji Watanabe[6], Takashi Taniguchi[7], Eli Rotenberg[2], Feng Wang[4], Johannes Lischner[3], Archana Raja[1]* and Alexander Weber-Bargioni[1]*

1-The Molecular Foundry, Lawrence Berkeley National Laboratory, Berkeley, USA.
2-Advanced Light Source, Lawrence Berkeley National Laboratory, Berkeley, USA.
3- Imperial College London, South Kensington Campus, London, United Kingdom.
4- Department of Physics, University of California at Berkeley, Berkeley, CA, USA.
5- Department of Physics, University of California at Santa Barbara, Santa Barbara, CA, USA
6- Research Center for Functional Materials, National Institute for Materials Science, 1-1 Namiki, Tsukuba, Japan 8
7-International Center for Materials Nanoarchitectonics, National Institute for Materials Science, 1-1 Namiki, Tsukuba,Japan
8-These authors contributed equally

‡-Current affiliation: Center for Nanotechnology Innovation @ NEST, Instituto Italiano di Tecnologia, Pisa,Italy
*antonio.rossi@iit.it, i.maity@imperial.ac.uk, araja@lbl.gov, awb@lbl.gov,

## Sample Characterization

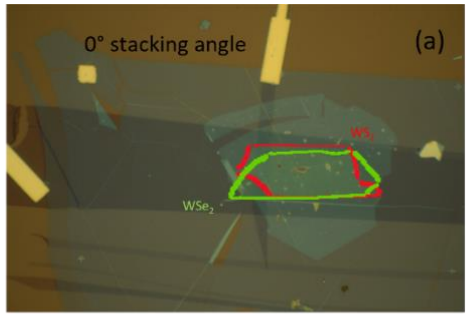 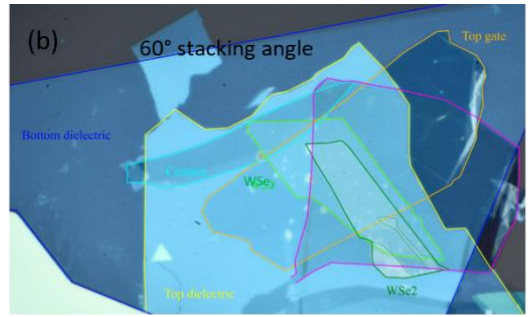

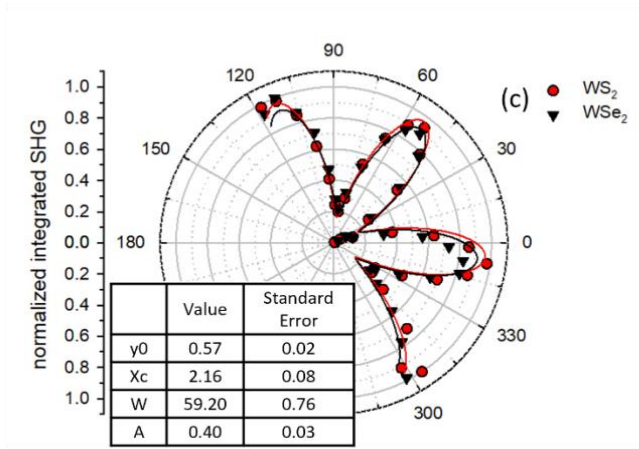 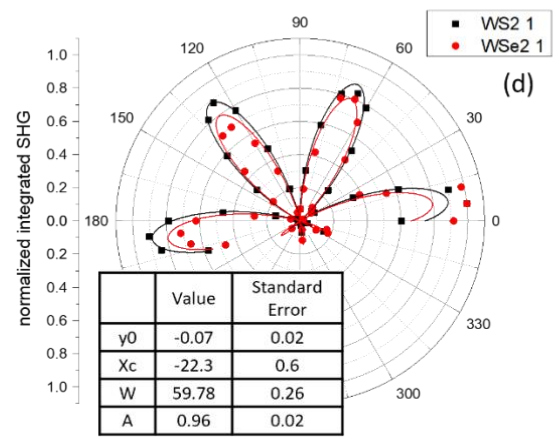

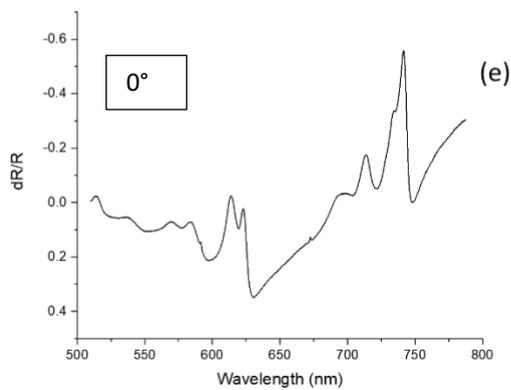 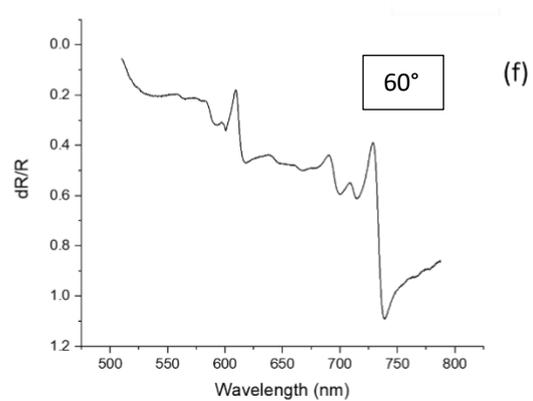

Fig S 1 Optical image of the samples used for this work. (a) and (b) are images of the 0° and 60° stacking angle samples respectively. (c-d) Second Harmonic Generation (SHG) analysis confirming the relative orientation of WS2 and WSe2 monolayers to be 0° and 60 respectively. (e-f) Adsorption spectra for the two samples. The sample in panel (a) is the same sample used in Ref. [1].

## PL spectrum multipeak fitting

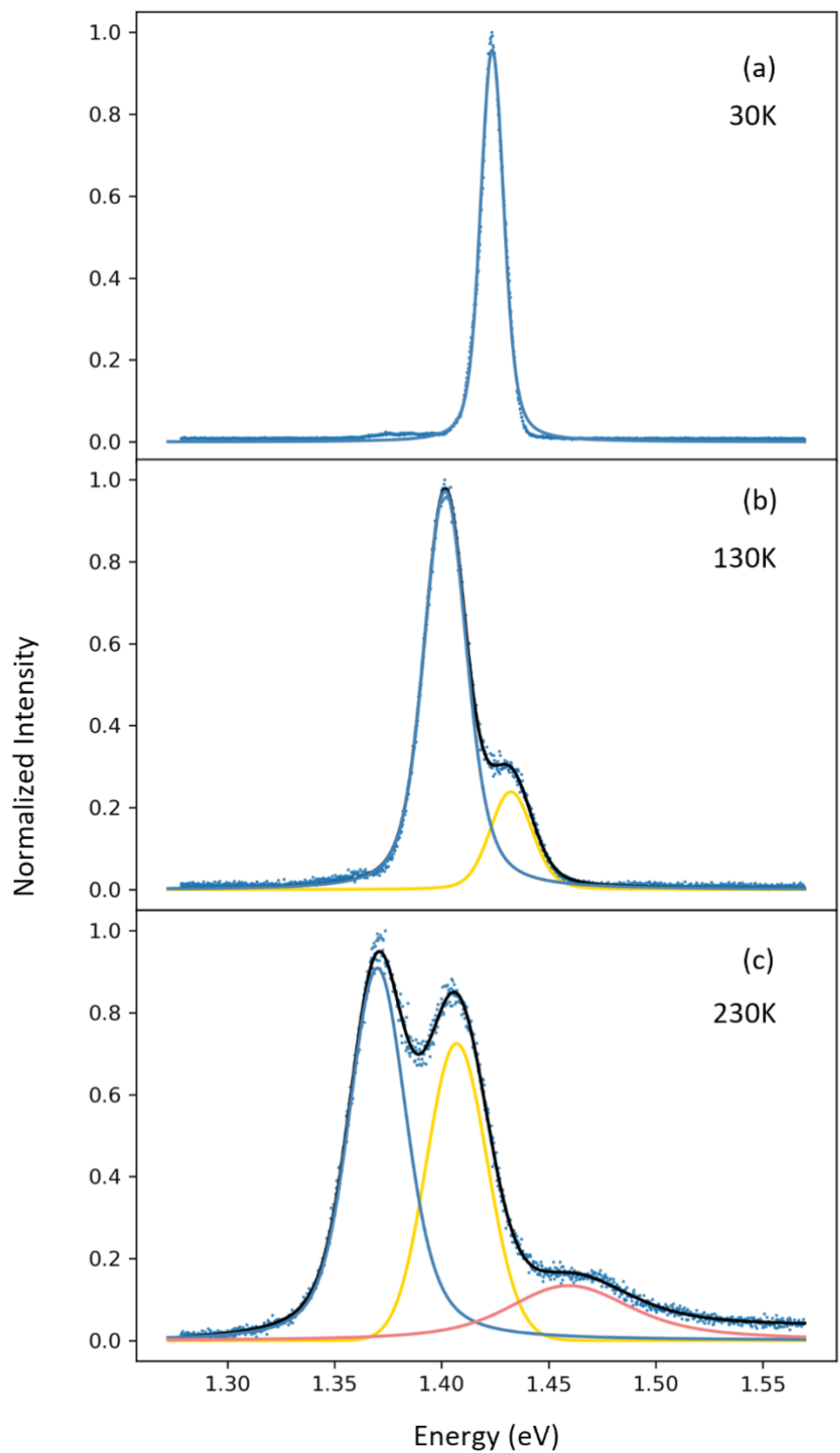

*Fig S 2 PL peak fitting for 0° sample at (a) 30 K, (b) 130K and (c) 230 K. The fitting is performed with one, two and three Voigt functions respectively.*

## Diffusion Measurements

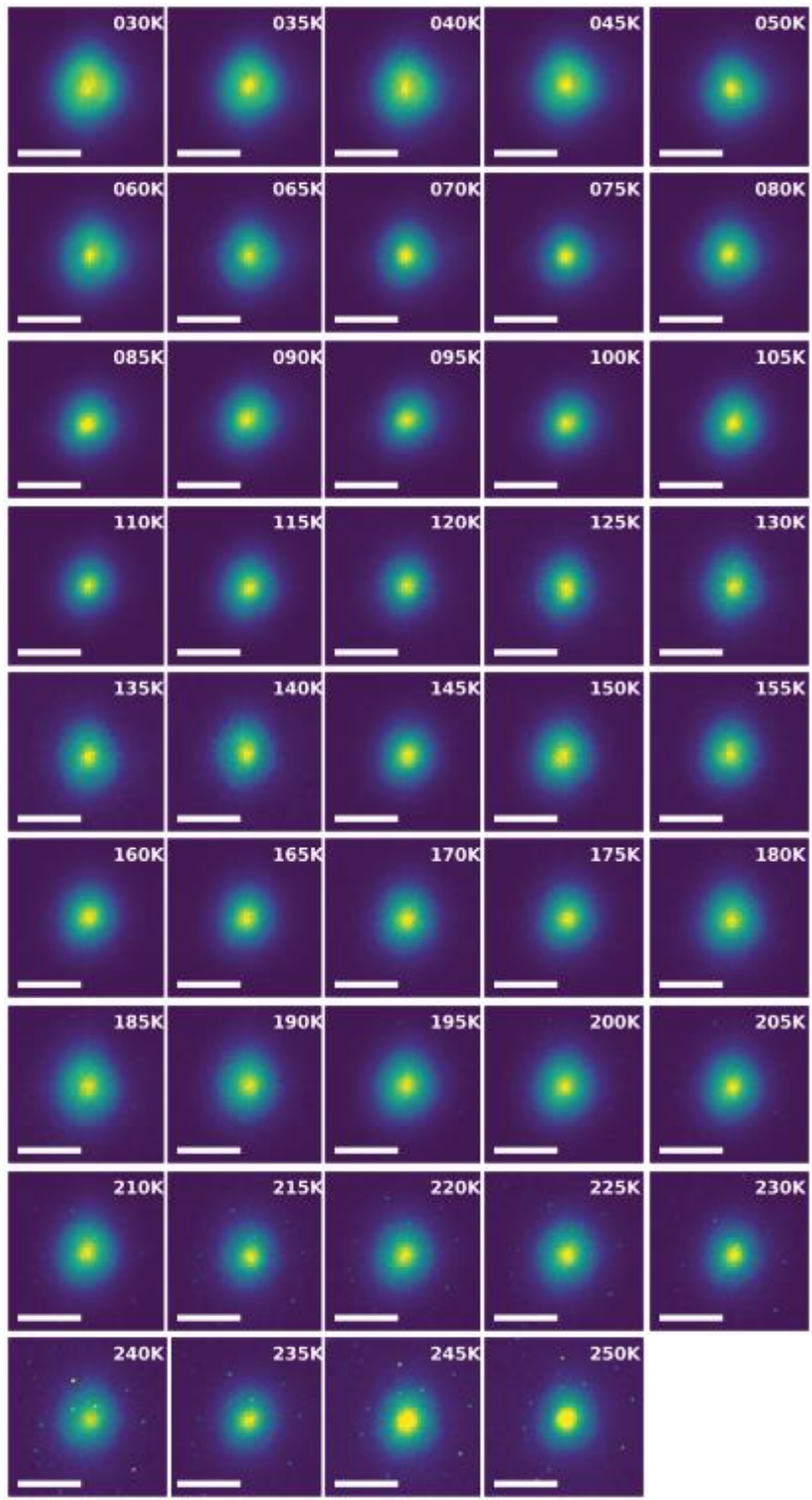

Fig S 3 Spatially resolved PL emission from 0° stacking angle sample. Each image is taken at different temperatures, ranging from 30 K to 250 K.

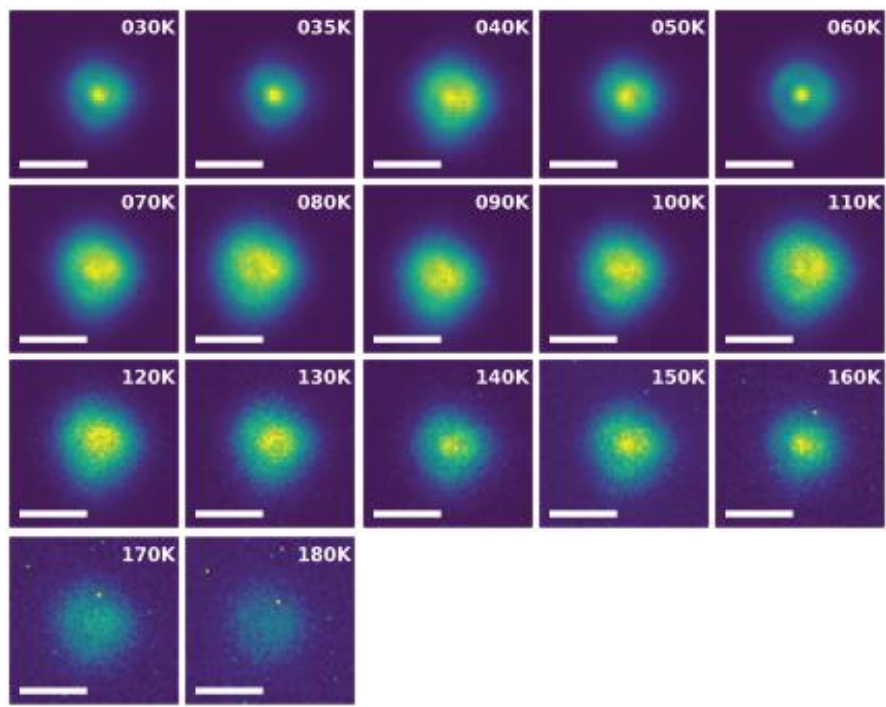

*Fig S 4 Spatially resolved PL emission from 0° stacking angle sample. Each image is taken at different temperatures, ranging from 30 K to 180 K*

# Exciton Diffusion single layer

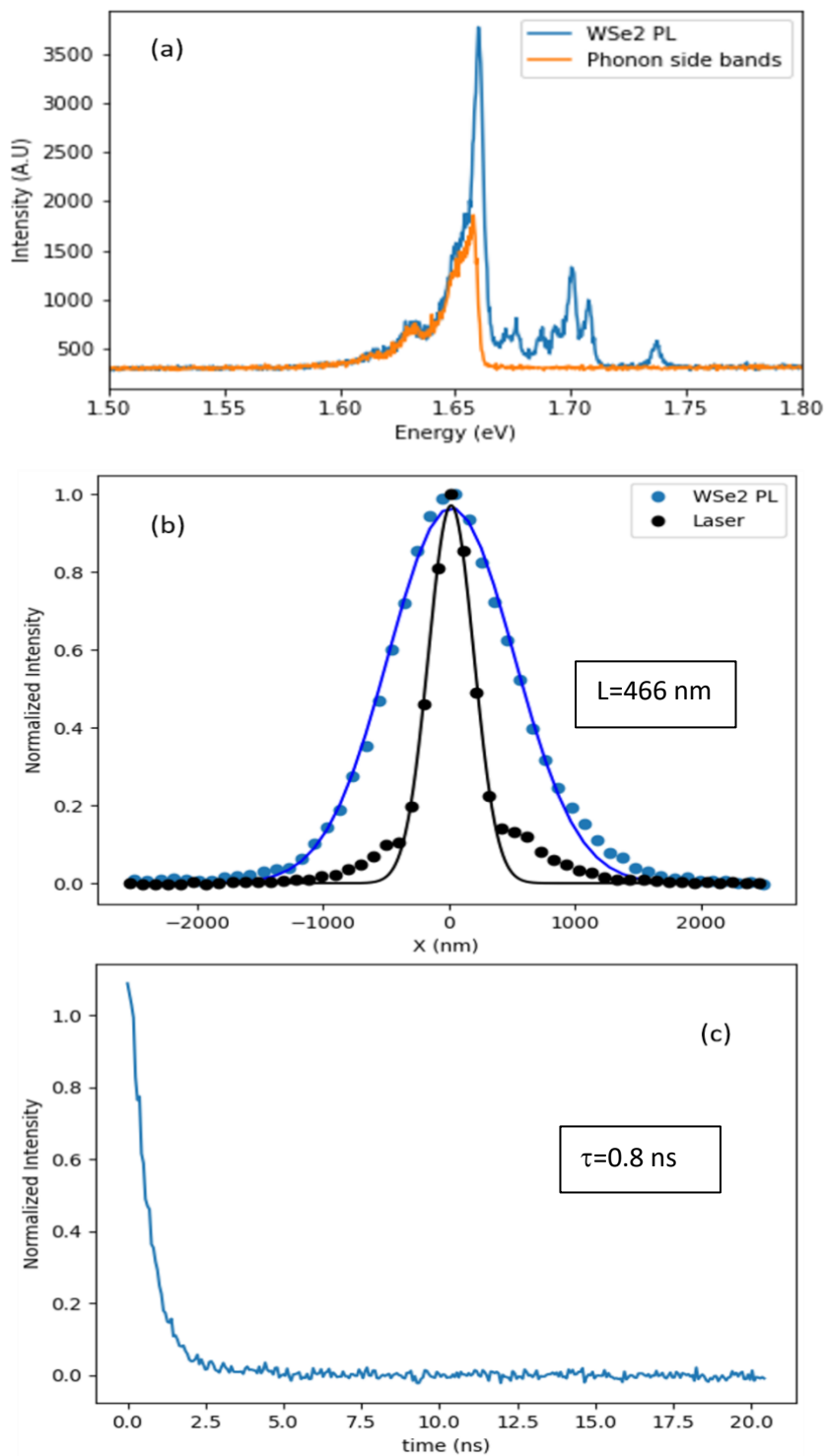

Fig S 5 To estimate the exciton diffusivity at cryogenic temperature in monolayer we measure the dark exciton dynamic through their spectrally well separated phonon side-bands (PSB)[2]. (a)PL spectra of WSe2 (blue) and PSB only (orange). Spatially(b) and (c) time resolved PL from PSB of isolated $WSe_2$ crystal The extracted diffusion length and lifetime are reported in the insets. The calculated Diffusivity is D = 1.35 $cm^2$/s

# Additional details on DFT Calculations

### A. Moiré potential using unit-cell calculations

The unit-cell lattice-constants of the $WS_2$ and $WSe_2$ are 3.18 and 3.32 Å, respectively. The lattice mismatch (about 4%) gives rise to a moiré when $WS_2$ and $WSe_2$ are stacked together even without a twist angle. We extract the moiré potential using the approach discussed in the main text. We also strain both the layer (isotropic tensile strain for $WS_2$ and compressive strain for $WSe_2$) and set the unit-cell lattice to 3.25 Å. We find that the qualitative moiré potential differences between 0° and 60° stacking are similar to experiments (see Fig. S6). However, these unit-cell calculations that do not include atomic reconstructions hugely underestimate the moiré potential for exciton hopping (which were shown in the main text). All the calculations are performed with the Quantum ESPRESSO package [3].

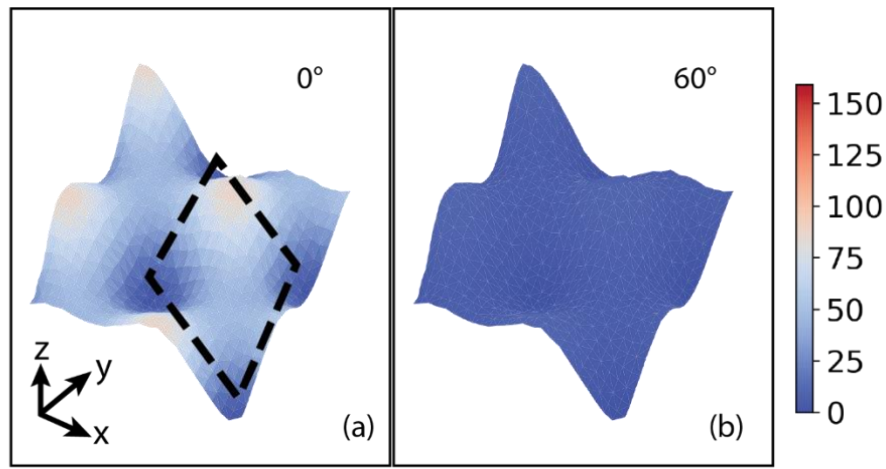

*Fig S 6. Moiré potential for the 0° and 60° alignment evaluated without atomic reconstruction.*

### B. Moiré potential extraction from twisted bilayer

We develop a simple scheme to compute the moiré potential using large-scale DFT calculations by taking into account the full atomic reconstructions. In this approach, we explicitly consider the localization of the electronic wave functions and extract the local band gap for different high symmetry stacking regions. To illustrate this, we plot the electronic wave function $|\psi_\Gamma|^2$ for several bands near the band gap (see Fig. S 7). The valence bands are marked as V1, V2,... . V1 corresponds to Valence Band Maximum (VBM). Similarly, the conduction bands are marked as C1, C2,... . C1 corresponds to Conduction Band Minimum (CBM). The local band gap at the red diamond stacking is 0.775 eV and obtained from V2 and C1 bands. The local band gap at the orange diamond stacking is 0.843 eV and obtained from V1 and C4 bands. The local band gap at the green diamond stacking is 0.963 eV and obtained from V3 and C4 bands. Similarly, we obtain the local band-gaps for 0° twisted $WS_2/WSe_2$ heterobilayer.

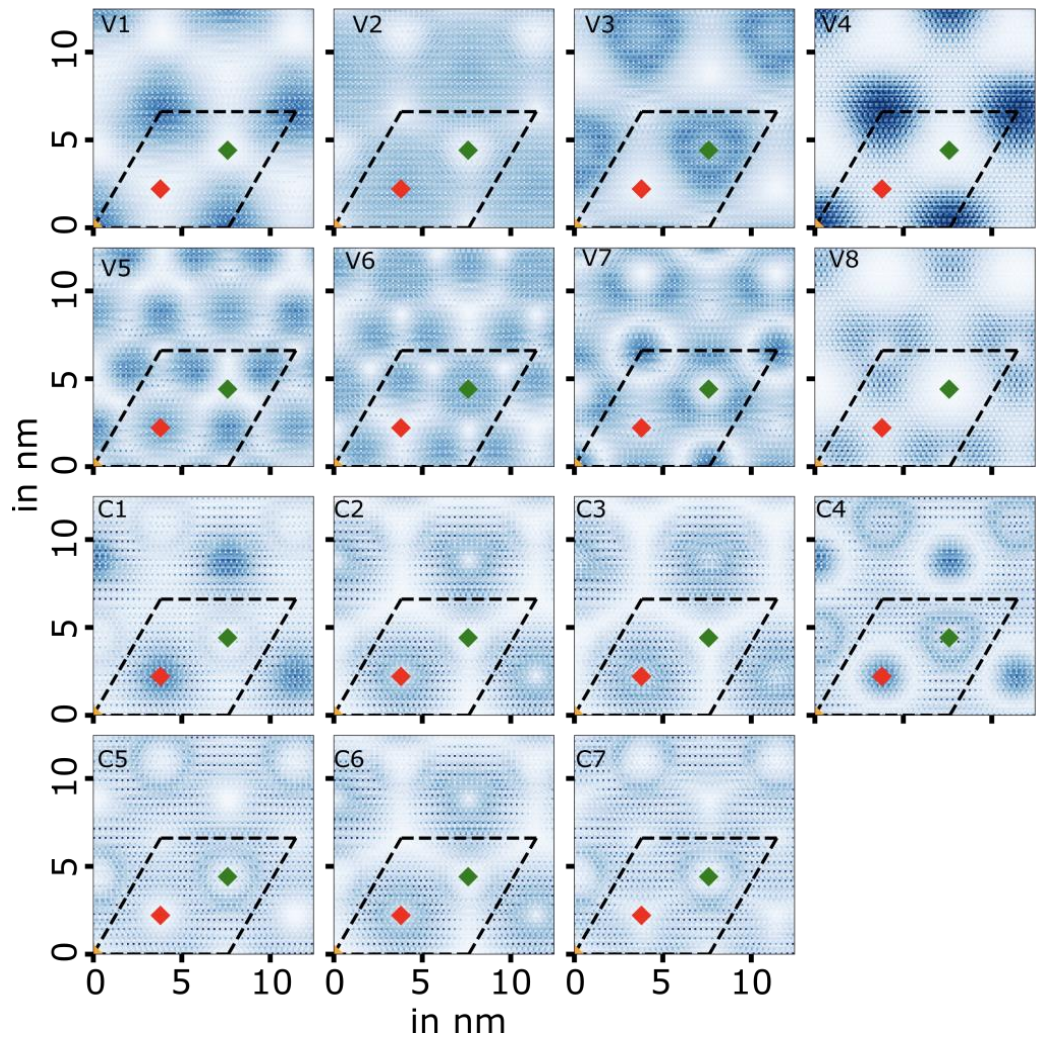

Fig S 7. $|\psi_\Gamma|^2$ averaged in the out-of-plane directions for several bands of 60° twisted $WS_2/WSe_2$. The unit cell is marked with dashed lines and different high-symmetry stackings are marked with coloured diamonds.

## C. Moiré movements at different times

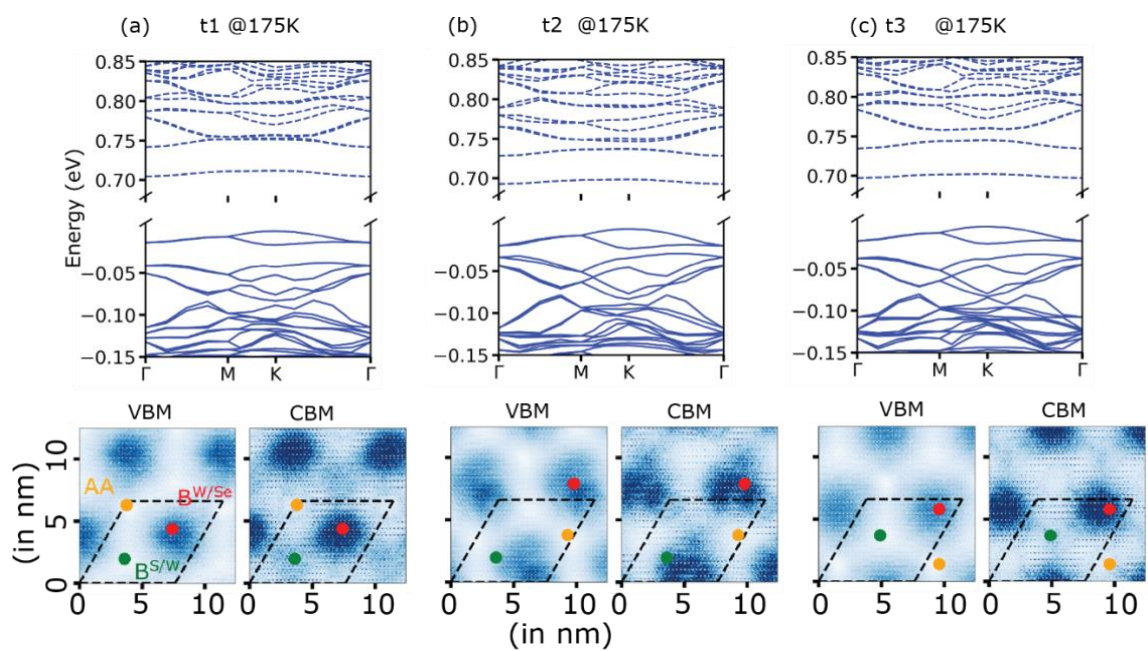

Fig S 8 Electronic band-structures using density functional theory of the structures obtained from classical molecular dynamics simulations in the canonical ensemble at 175 K. The $|\psi_\Gamma|^2$ averaged in the out-of-plane directions for the valence band maximum (VBM) and conduction band minimum (CBM) are shown in the bottom panel for the same structures. The center of different high-symmetry stackings in the moir´e pattern are also marked.